# Radio-Frequency-Driven Reshaping of the Mesoscale Charge-Density-Wave Landscape in 1T-TaS$_2$ Thin-Film Devices

Maedeh Taheri[1,2,3], Zahra Ebrahim Nataj[1,2,3], Nick Sesing[4], Topojit Debnath[5], Tina T. Salguero[4], Roger K. Lake[5,*], and Alexander A. Balandin[1,2,3*]

[1]Phonon Optimized Engineered Materials Laboratory, Department of Materials Science and Engineering, University of California, Los Angeles, California 90095 USA

[2]California NanoSystems Institute, University of California, Los Angeles, California 90095 USA

[3]Center for Quantum Science and Engineering, University of California, Los Angeles, California 90095 USA

[4]Department of Chemistry, University of Georgia, Athens, Georgia, 30602 USA

[5]Laboratory for Terahertz and Terascale Electronics, Department of Electrical and Computer Engineering, University of California, Riverside, California 92521 USA

---

* Corresponding authors: balandin@seas.ucla.edu; rlake@ucr.edu

# Abstract

Radio-frequency excitation directly reshapes the mesoscale charge-density-wave landscape in quasi-two-dimensional 1T-TaS$_2$ thin films. Under combined RF and DC bias, the hysteretic current–voltage characteristics associated with the nearly commensurate–incommensurate transition are strongly altered, displaying RF-driven collapse, branching, and multiple step-like features that depend on frequency and drive amplitude. *In-situ* Raman measurements show enhanced intensity and linewidth narrowing of low-frequency CDW phonon modes, consistent with reduced dephasing and increased coherence of the periodic lattice distortion under RF drive. This behavior is captured by combining an overdamped time-dependent Ginzburg–Landau description of the commensurate CDW with a morphology-informed percolative resistor-capacitor transport model. The simulations indicate that oscillatory driving anneals frustrated domain configurations, reduces domain-wall density, and reorganizes the discommensuration network, while the transport model reproduces the resulting hysteresis, avalanche-like pathways, and RF-induced conductance steps. RF driving therefore provides an effective route for controlling collective electron–phonon order and accessing metastable transport states in 1T-TaS$_2$, with implications for reconfigurable RF electronics, memory, and unconventional computing based on correlated materials.

1. Introduction

Correlated electronic systems can host a variety of quantum states, such as charge-density waves (CDW), superconductivity, and Mott insulating phases [1-6]. These electronic phases often compete, and their ground states can be tuned by external stimuli, including temperature, pressure, strain, and doping [7-11]. 1T-TaS$_2$, a layered quasi two-dimensional (2D) material, can provide a model platform for exploring quantum order in correlated materials, owing to its rich array of emergent quantum phases and strong nonlinearity [12-18]. Switching between different electrical resistance states has been achieved through both thermal and non-thermal mechanisms [21-26]. Non-equilibrium excitation with ultrafast optical or THz pulses has been shown to destabilize the ground state and promote the formation of metastable states [27-31]. In this technique, the photoinduced carrier density can modulate the electron-electron and electron-phonon interaction near the Fermi level, suggesting a carrier-doping-driven mechanism, rather than a thermally driven pathway [29]. Further studies suggest a charge-configuration memory 1T-TaS$_2$ device, based on a non-thermal "writing" mechanism using a picoseconds electric pulse, followed by a thermal "erasing" process [32]. Injecting carriers through electrodes in the commensurate state gave rise to the metastable textured domain states [22, 33, 34]. While numerous quantum phases have been revealed through ultrafast measurement, the response of 1T-TaS$_2$ to combined AC and DC excitation remains largely unexplored. The interplay of two competing periodicities has been shown to produce Shapiro-like anomalies in the *I-V* characteristic of quasi-1D materials [1, 35, 36]. Although the AC response of a driven CDW system has been a focus of extensive research in quasi-1D materials, it has yet to be addressed in quasi-2D systems. Here in this work, we investigate the response of quasi-2D 1T-TaS$_2$ under medium and high–range radiofrequency (RF) excitation, exploring AC-DC coupled behavior across different temperature ranges.

The 1T-TaS$_2$ crystals and thin films undergo a succession of phase transitions as a function of temperature under equilibrium conditions. At high temperature, $T < 550$ K, the crystal undergoes a metallic to incommensurate CDW transition. An incommensurate to nearly commensurate (IC–NC) CDW transition occurs with a hysteresis window at $T \approx 334 - 356$ K. At this transition, the incommensurate charge-density-wave makes a rotation of ~11º–12 º relative to the crystal axis to create a NC–CDW state, which is accompanied by crystal lattice distortion and reconstruction in the commensurate regions [12, 37]. In this state, which is stable at room temperature, hexagonally ordered commensurate domains form star-of-David clusters. The domains are separated by discommensuration or border walls, where the CDW undergoes a phase slip and result in compression or expansion of wavelength [37]. The switch from the nearly commensurate to commensurate (NC–C) CDW state happens with a hysteretic first-order transition at $T \approx 130 - 220$ K. In the C–CDW, the domains evolve into a long-range uniform commensurate state, in which the CDW wavevector is oriented at an angle of ~13.9º relative to the lattice. Upon heating the sample to $T \approx 220$ K, the crystal transitions into a triclinic (T) phase which persists up to $T \approx 280$ K, before returning into an NC phase. In the T phase, the CDW loses the hexagonal symmetry and adopts a stripe-like domain structure [37, 38].

2. Results and Discussion

Single crystals of the 1T polytype of $TaS_2$ were grown using the chemical vapor transport (CVT) technique (see Supplementary Figure S1 for more details on crystal characterization). Few-layer 1T-$TaS_2$, ranging from ~10 to 35 nm, was mechanically exfoliated onto a Si/$SiO_2$ substrate. Thin films of selected 1T-$TaS_2$ were capped with hexagonal boron nitride (*h*-BN) for protection from environmental exposure. The contacts were patterned using electron beam lithography, with channel length ranging from ~1 to 3 $\mu$m, followed by dry etch of the *h*-BN layer and Ti/Au metal deposition. Figure 1(a) shows the circuit configuration of the 1T-$TaS_2$ device driven by a combination of both AC and DC. The RF signal in the range of 0.15 to 300 MHz is coupled with the DC supply, using Bias-Tee components, and applied to the device. Figure 1 (b) depicts the Scanning Electron Microscopy (SEM) image of a prototype 1T-$TaS_2$ device. The DC resistance of the 1T-$TaS_2$ device as a function of temperature shown in panel (c) confirms the quality of the sample and shows the thermodynamic transition with two hysteresis windows, defining the IC–NC transition at $T \approx 334 - 356$ K and NC–C at $T \approx 130 - 220$ K. Upon warming, 1T-$TaS_2$ enters the triclinic (T) phase which remains stable up to $T \approx 280$ K, before reverting to the NC phase. This is observed with a small kink in the resistance during the warming cycle, at $T \approx 280$ K. The arrows indicate the structures associated with the corresponding CDW phase.

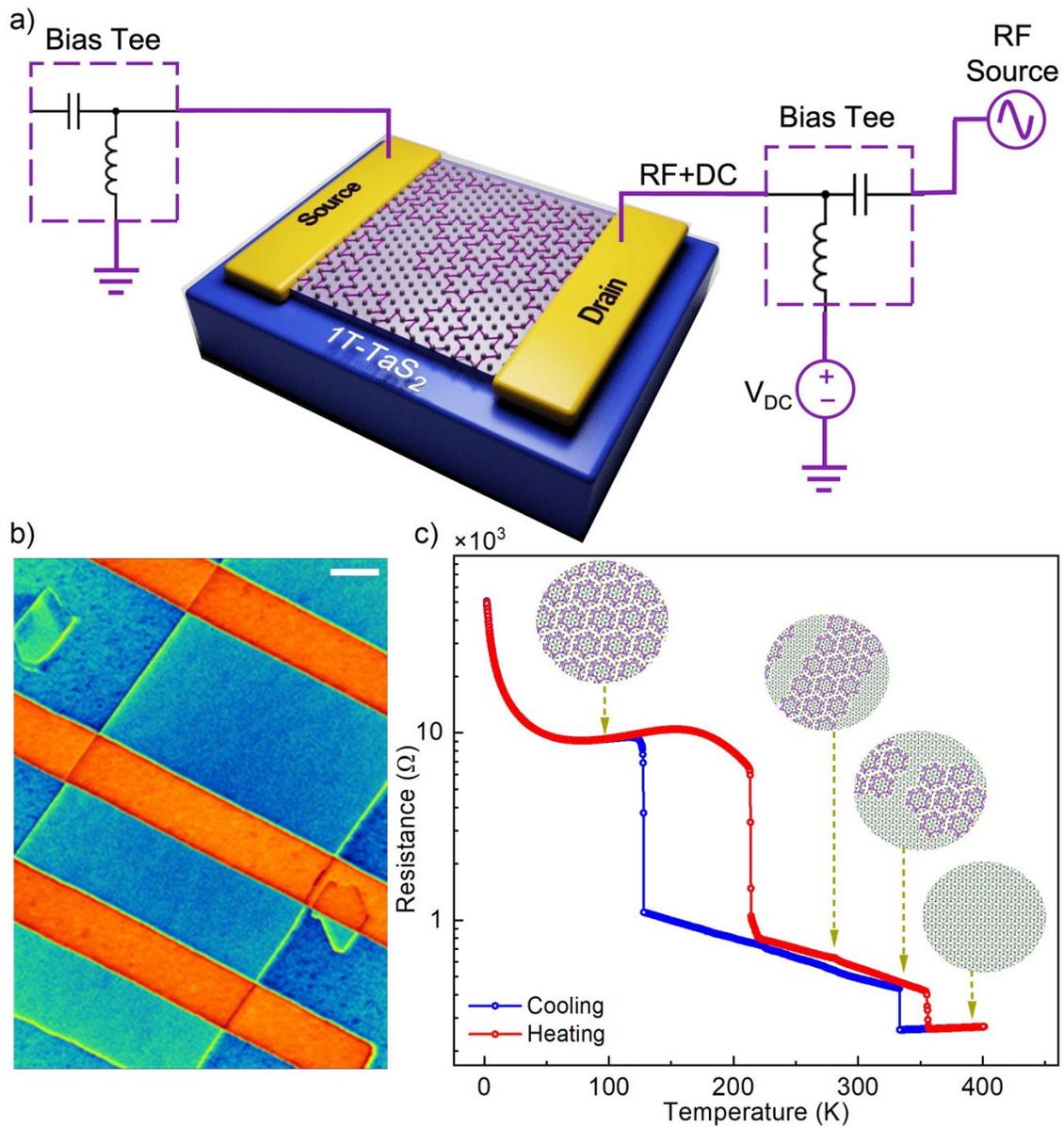

**Figure 1: DC characterizations and AC-DC circuit setup.** The circuit diagram of AC-DC coupled measurement can be seen in panel (a), where two Bias Tees isolate the RF and DC pathway. b) The SEM image of 1T-TaS$_2$ channel. The pseudo-coloring is used for clarity. The scale bar shows 1 $\mu$m. c) The temperature-dependent resistance measurement of 1T-TaS$_2$ device shows the IC-NC and NC to C-CDW transition, with two hysteresis window at $T \approx 334 - 356$ K and $T \approx 130 - 220$ K, respectively. The resistance is measured at low DC bias, 15 mV. The arrows show the illustration of crystal and formation of star-of-David domains, upon sweeping across the temperatures.

Nonlinear switching behavior in 1T-TaS$_2$ shares similarities with switching CDWs observed in quasi-1D crystals such as NbSe$_3$, TaS$_3$, and K$_{0.3}$MoO$_3$ [39, 40]. While conventional quasi-1D crystals show CDW depinning with a kink in *I-V*, there are selected crystals displaying an abrupt and hysteretic discontinuity, which is attributed to spatially nonuniform CDW pinning within the crystal. These samples are characterized by bistability in the *I-V*, negative differential resistance (NDR), large-amplitude 1/*f* noise, and inductive ac response [40, 41]. In systems with gradual depinning, the application of both AC and DC drives gives rise to the Shapiro-like steps and mode-locking phenomena [35, 36]. The response of switching samples to the AC-DC drive is more complex and suggests unusual Shapiro steps and a period doubling route to chaos [40, 41]. Instabilities observed in switching crystals can be categorized into two types, depending on whether they are driven by low-frequency or high-frequency excitation. Under low-frequency drive (*f* ≤ 1 MHz), the sample is repeatedly driven through the switching region of the DC *I-V* response, generating a characteristic power spectral signature, referred to as "AC switching noise". The time scale for the switching of quasi-1D samples is reported to be in the range of ~ 1 μs [40]. For frequencies above ~ 5 MHz, the system's response follows a different mechanism, in which the phase slip centers synchronize, resulting in a mode-locked steps in response to the external RF signal, i.e., the Shapiro-like steps.

To better understand the dynamics of quasi-two-dimensional 1T-TaS$_2$ under AC-DC excitation, we conducted low-frequency measurements, from 0.15 to 1.2 MHz, and high-frequency measurements, from 10 to 300 MHz. As it has been studied before [12, 13], the NC–IC CDW transition in 1T-TaS$_2$ can be induced with DC bias with a signature hysteresis window. Figures 2 (a) and (b) represent the *I-V* characteristics under *f* = 1 MHz radiation for devices at *T* = 295 K and 240 K, respectively, where the 1T-TaS$_2$ has an NC–CDW ground state. The bottom traces, labeled "No RF," show the hysteresis attributed to the NC–IC CDW transition, with a single-step switching event. Upon increasing the RF power at *f* = 1 MHz radiation, the hysteresis window deforms and splits into two smaller hysteresis windows. The onset current of the second hysteresis, labeled as $I_{H2}$, scales almost linearly with frequency from ~0.15 to 1.2 MHz, as can be seen in Figure 2 (c) for two devices. The power spectrum of the 1T-TaS$_2$ output in the AC-DC driven state with *f* = 1 MHz and $V_{H1} - V_{ac} < V_{dc} < V_{H1} + V_{ac}$ is shown in Figure 2 (d) at *T* = 260 K, where $V_{H1}$ is the onset voltage of the first hysteresis (note Figure 2(d) inset). The frequencies

6 | P a g e

up to the fifteen harmonics (*f* = 15 MHz) can be seen in the spectrum due to the nonlinearity of the switching region. A similar form of noise was reported in the power spectrum of 1D switching samples – AC switching noise – which emerges when a sample is repeatedly cycled through switching in the DC *I-V* curve with low-frequency drive [40]. The spectrum exhibits a broadband noise superimposed on sharp peaks at the driving frequency and its harmonics. In the 1D switching samples, the AC switching noise was attributed to the avalanche-like depinning of domains from ultra-strong pinning centers, which led to the appearance of harmonics and broadband components [40].

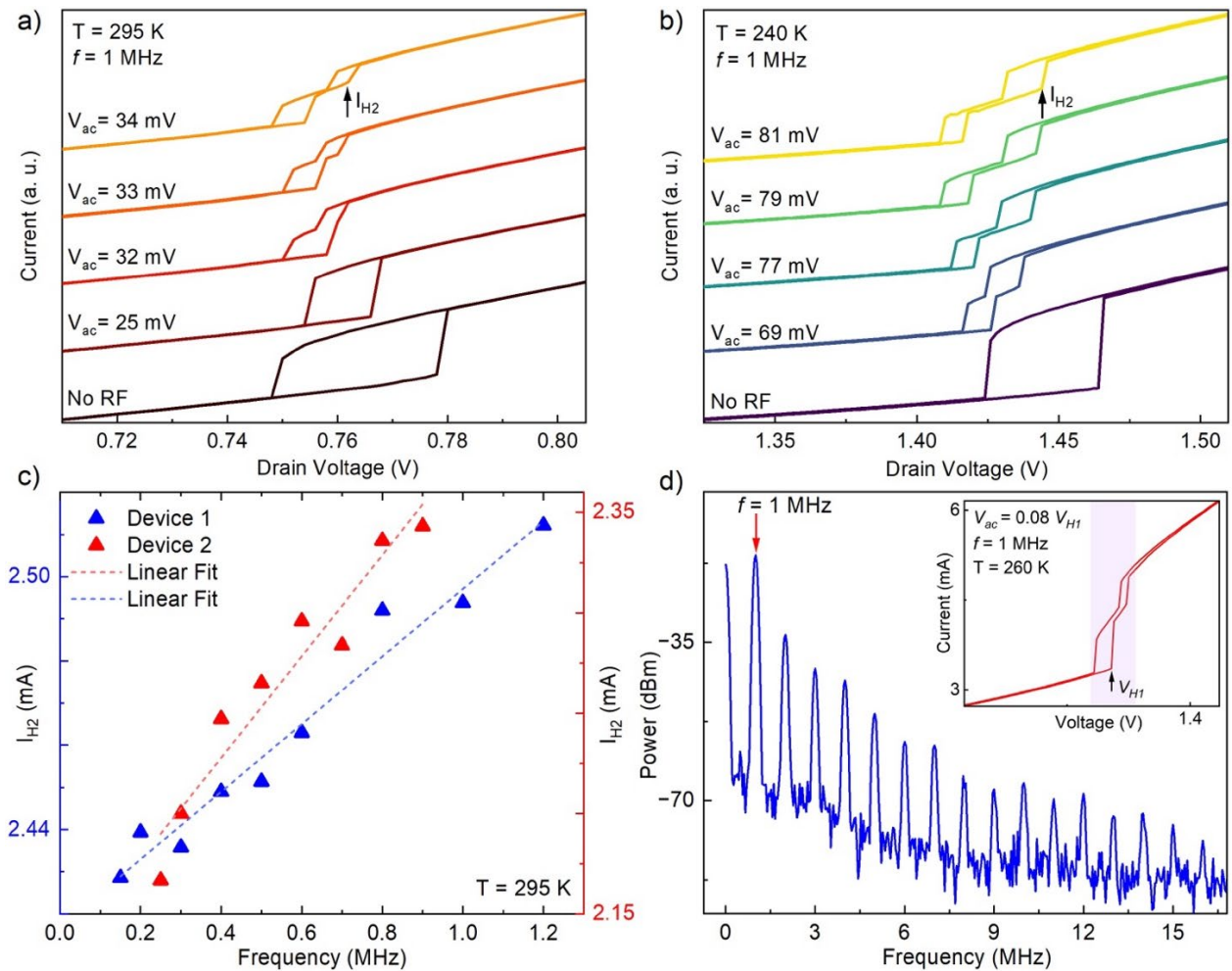

**Figure 2: Low-frequency AC-DC response in NC-IC CDW regime .** The evolution of hysteresis is demonstrated in AC-DC measurment, with $f$ = 1 MHz, for devices at two different temperatures, $T$ = 295 K (a) and $T$ = 240 K (b). The shifts in the seperation point, the onset current of second hysteresis $I_{H2}$, as a function of the drive frequency is plotted in panel (c) at room temperature for two devices. The current exhibits an approximately linear dependence on frequency from 0.15 to 1.2 MHz. (d) The output power spectrum of the device under RF radiation with $f$ = 1 MHz, shows the presence of harmonics due to the nonlinearty in the respsone. The pink shaded region in the inset, describes the DC bias points of the chnanel, where $V_{H1} - V_{ac} < V_{dc} < V_{H1} + V_{ac}$.

The response of the 1T-TaS$_2$ to the higher frequency drive, $f$ = 70 to 300 MHz is presented in Figure 3, in the NC-IC CDW regime. Two regions are highlighted: near the NC-IC CDW hysteresis (panel (a)), and within the NC state below the hysteresis window (panel (c)). Figure 3 (a) displays forward-sweep $I$-$V$ characteristics measured at $f$ = 70 MHz and $T$ = 270 K, focused on the vicinity of the hysteresis region for increasing RF amplitudes. Compared to the No-RF trace, the application of RF excitation shifts the NC-IC transition to lower drain voltages and progressively modifies the transition region, where additional current steps emerge (blue arrows). Panel (b) shows the differential conductance as a function of RF power in the vicinity of the NC-IC CDW regime, under $f$ = 70 MHz radiation. With increasing RF power, the NC-IC step shifts to lower bias, and additional conductance steps appear (see blue lines starting at drain voltage of ~1.3 V and $P_{RF}$ = -2 dBm). Further information on the differential resistance evolution and frequency dependency of the steps from $f$ = 10 to 150 MHz, are available in Figure S2 of the Supplemental Material. Distinct resistance anomalies appear sequentially with increasing $V_{ac}$: the first step becomes visible near $V_{ac} \approx 155$ mV, a second step develops around $V_{ac} \approx 195$ mV, and finer features – manifested as smaller peaks in $dV/dI$ – emerge at higher RF amplitudes. Upon cooling the device to $T$ = 250 K (see Fig. 3(c)), the measured $I$-$V$ characteristics under $f$ = 70 MHz radiation exhibit a series of hysteretic steps that appear within the NC state, before the NC–IC transition (pink shaded region).

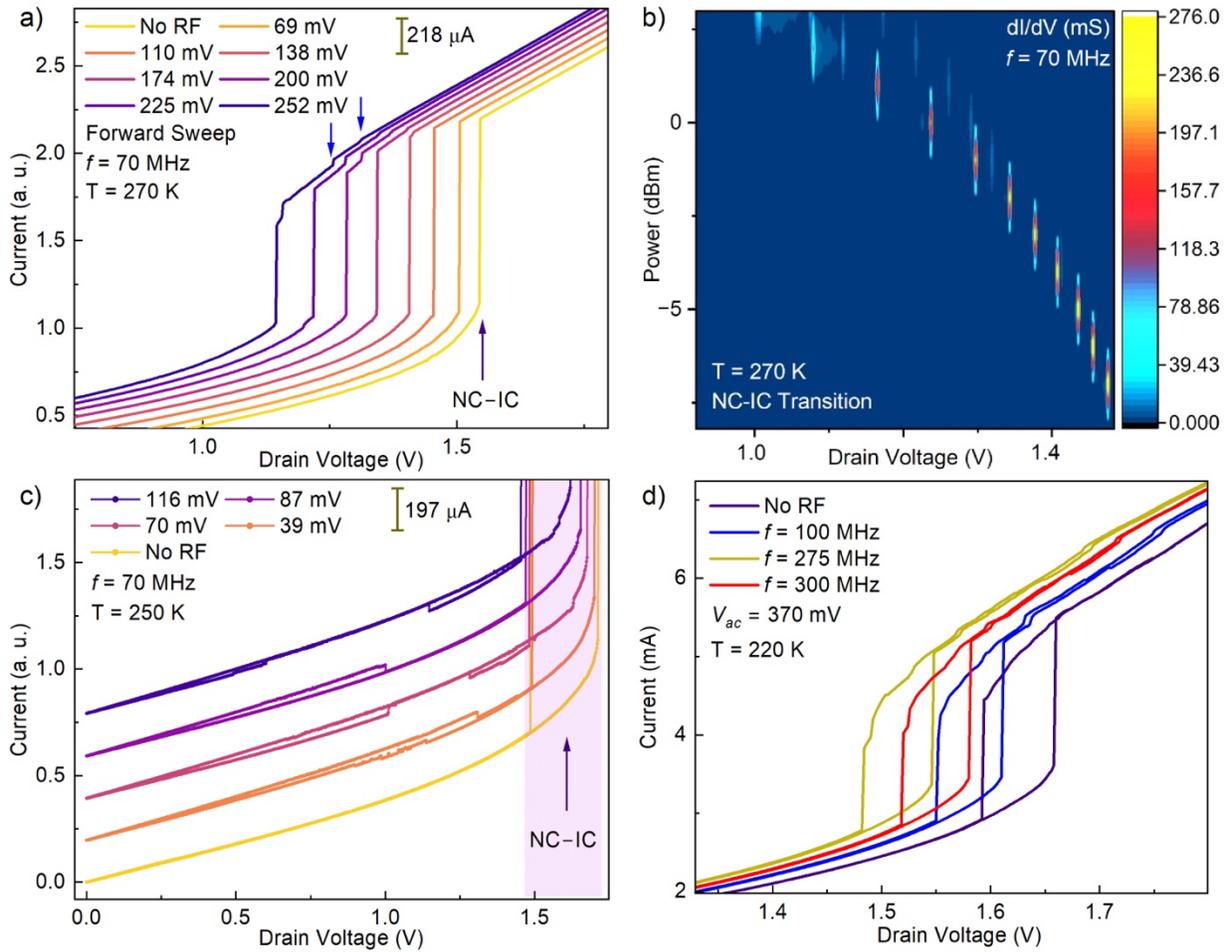

**Figure 3: High frequency AC-DC response in NC-IC CDW regime.** Panel (a) shows the *I-V* under 70 MHz radiation in forward sweep at *T* = 270 K. The "No RF" trace shows the NC-IC transition. Upon increasing the AC amplitude, the onset of transition reduces, and extra steps are separated as it is shown with blue arrows. b) The color map shows the measured differential conductance as a function of RF power, in the vicinity of NC-IC transition hysteresis at *T* = 270 K. As the RF power increases, the NC-IC step moves to the lower biases and more conductance steps emerge. c) Measured *I-V* curves at *T* = 250 K under *f* = 70 MHz excitation show the emergence of hysteretic steps at lower bias within NC state, preceding the onset of NC-IC regime (the pink shaded region). Panel (d) shows frequency dependent *I-V*s measured under RF excitation from 100 to 300 MHz.

In the NC state, the CDW in 1T-TaS$_2$ forms a mosaic of domains separated by a diffuse and irregular discommensuration network, as revealed by STM studies [34, 42]. At the local scale, the elastic response of the CDW can be disrupted through plastic deformations of this network [43, 44]. Such deformations may be topologically trivial—such as domain-wall rearrangements—yet can give rise to metastable configurations associated with multiple local energy minima. In the low-bias regime of the NC state, where Joule heating progressively reduces the domain size, RF excitation can couple to the discommensuration network and dynamically modulate domain-wall configurations [45]. Here, the discrete hysteretic switching branches, appearing in the NC state, can be seen in Figure 3(c). These branches do not evolve monotonically with RF amplitude and instead occur sporadically over a limited bias range, suggesting the involvement of localized, metastable CDW configurations rather than a global phase transformation. Importantly, once formed, these hysteretic branches persist upon removal of the RF excitation and remain observable in subsequent DC sweeps, indicating that RF driving possibly stabilizes long-lived metastable states within the NC phase.

Figure 3(d) shows frequency-dependent AC–DC measurements performed from $f$ =100 to 300 MHz at $T$ = 220 K. Within this frequency range and at a fixed AC amplitude of $V_{ac}$ = 370 mV, the overall hysteresis area remains approximately constant, while the positions of the step features shift nonlinearly with frequency. Additional details on the frequency evolution of these steps are provided in Supplemental Figure S2 (b). Previous STM studies of 1T-TaS$_2$ have demonstrated that, in the NC state, the CDW is not fully phase-locked; instead, both the phase and amplitude are spatially nonuniform, evolving more slowly within domain interiors than along the surrounding discommensuration network [37]. As the DC bias increases and the system approaches the hysteretic regime in the *I-V* characteristics, the characteristic CDW domain size decreases, suggesting increased spatial variation of the CDW phase across the hysteresis window. In contrast to the Shapiro mechanism in quasi-one-dimensional CDW systems—where phase locking leads to a linear relationship between CDW current and frequency—the step features observed in 1T-TaS$_2$ exhibit a deviation from linear scaling as a function of frequency. Moreover, because the onset of sliding in quasi-two-dimensional CDW systems is not necessarily accompanied by pronounced nonlinear transport, isolating the CDW contribution to the total current remains challenging [46, 47]. Consistent with this picture, analysis of the output power spectrum in AC–DC measurements of 1T-TaS$_2$ does not reveal harmonics or signatures of period doubling en route to chaos. Taken together, these observations reduce the likelihood that the observed step features arise from partial phase locking to the external drive analogous to that in quasi-one-dimensional CDW systems. Similar to the small hysteretic steps observed within the NC state (Figure 3(c)), the step features appearing in the vicinity of the NC–IC CDW transition persist after the RF excitation is switched off. As the CDW domains progressively melt and the discommensuration network expands, the RF field can couple more effectively to domain walls, promoting their motion and inducing reorganization of the CDW domains boundaries.

To gain insight into the behavior of CDW domains and their interaction under AC–DC driving, we performed in-situ, bias-dependent Raman measurements under both DC and combined RF excitation. We first examine the evolution of the Raman-active CDW modes under DC bias alone to establish the correspondence between transport behavior and the C–IC CDW transition, before extending the measurements to RF-driven conditions. Figure 4(a) shows DC bias–dependent Raman spectra of 1T-TaS$_2$ measured in the commensurate CDW phase at $T = 100$ K, where well-defined low-frequency, zone-folded modes [48-50] below 100 cm$^{-1}$ are observed at zero bias. The light blue curve shows the Gaussian fit to the zone-folded mode at 60, 70, and 77 cm$^{-1}$. As the DC bias is increased, these modes progressively broaden and weaken, accompanied by a pronounced increase in their full width at half maximum (FWHM), signaling the evolution toward the incommensurate CDW state. The Raman intensity of the 77 cm$^{-1}$ mode, normalized to the silicon reference peak at each bias, is shown in Figure 4(b). The pronounced decrease in intensity from approximately 1.2 to 0.08 marks the transition from the commensurate to the incommensurate CDW state.

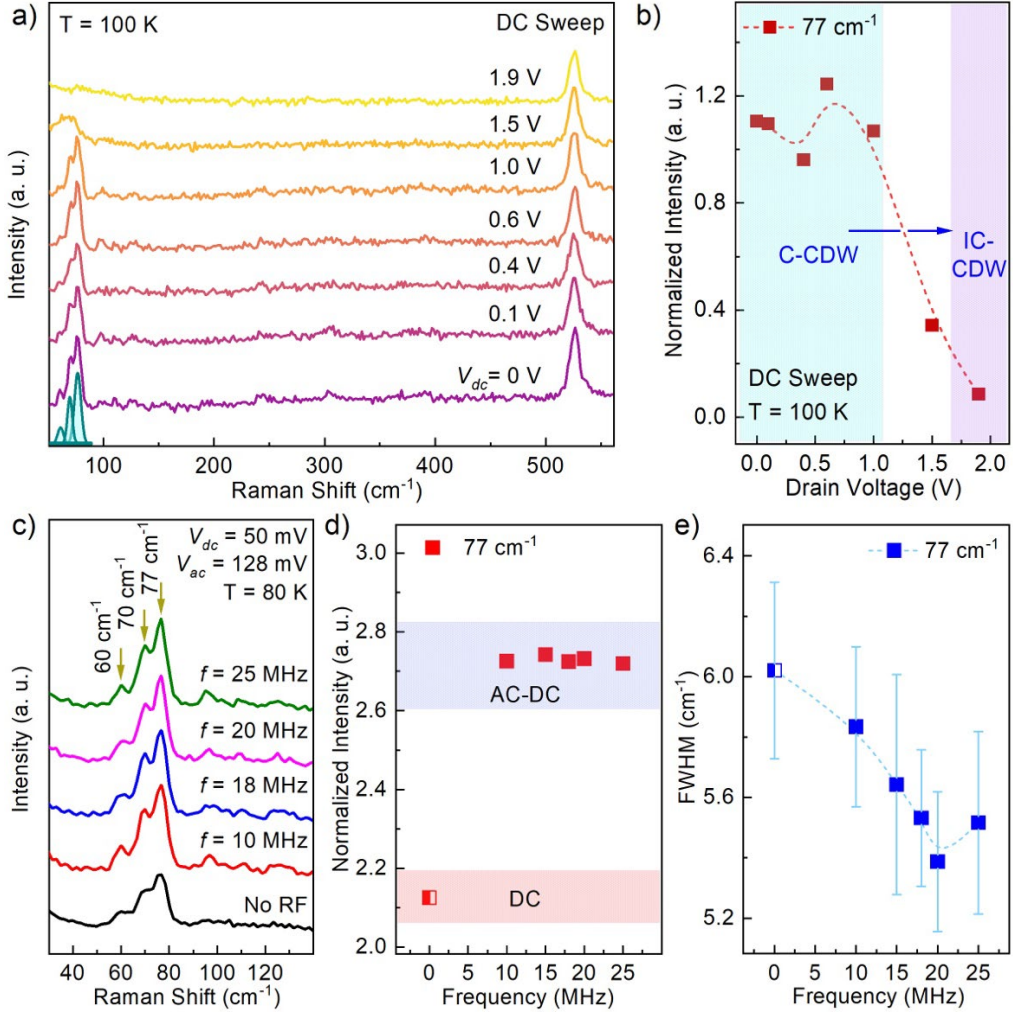

**Figure 4: In-situ bias-dependent Raman measurement.** a) The DC bias-dependent Raman spectra of 1T-TaS$_2$ channel is measured in the commensurate state at T = 100 K, with a 488-nm laser. At zero bias, the low-frequency modes can be seen in the spectral range < 100 cm$^{-1}$. As the DC bias, $V_{dc}$, is increased, the peaks broadens and the intensity decreases, where it enters the incommensurate state at around $V_{dc}$ = 1.9 V. The light blue curves represent the Gaussian fits to the vibrational modes at 60, 70, 77 cm$^{-1}$. Panel (b) shows the intensity of 77 cm$^{-1}$ peak, which is normalized to the silicon peak, as a function of DC bias voltage, where the channel makes the transition from C to IC-CDW. c) The Raman spectra of the channel under the AC-DC drive is measured in the commensurate state for range of frequencies from $f$ = 10 to 25 MHz, where $V_{dc}$ = 50 mV and $V_{ac}$ = 128 mV and compared against the spectrum in the absence of RF. Panel (d) and (e) shows the normalized intensity and FWHM of the 77 cm$^{-1}$ mode as a function of frequency, respectively.

To isolate the effect of RF excitation on the CDW lattice dynamics, Raman spectra were subsequently acquired under simultaneous DC and RF bias ($V_{dc}$ = 50 mV, $V_{ac}$ = 128 mV), as shown in Figure 4(c), over a frequency range of $f$ = 10 to 25 MHz at $T$ = 80 K. Focusing on the 30 – 140 cm$^{-1}$ spectral range under AC-DC excitation and comparing it to the "No RF" spectrum (black curve), the peaks are observed to become sharper. The normalized intensity of the 77 cm$^{-1}$ mode is shown in Figure 4(d). Relative to the DC measurements, RF excitation produces a measurable enhancement of the Raman signal. At the same time, the linewidth of 77 cm$^{-1}$ mode shows a decrease under RF excitation, as plotted in Figure 4 (e). Since these low-frequency modes originate from the periodic lattice distortion (PLD) associated with the commensurate CDW, their Raman intensity reflects the spatially averaged amplitude and coherence of the PLD, and their FWHM is sensitive to dephasing and inhomogeneous broadening arising from domain walls, local strain, and phase fluctuations [37, 49, 51]. The increased intensity and decreased FWHM indicate greater spatial homogeneity and phase coherence of the CDW under RF drive.

To gain further insight into the dynamics of the commensurate CDW and its domain-wall network, we modeled the nonequilibrium evolution of the order parameter using overdamped time-dependent Ginzburg–Landau (TDGL) [52] dynamics [Figures 5(a)–(f)]. We adopt a coarse-grained two-dimensional description of the commensurate CDW in terms of a complex order parameter $\psi(r,t) = A(r,t)e^{i\phi(r,t)}$, where $A(r,t)$ denotes the CDW amplitude and $\phi(r,t)$ the phase [44, 52, 53]. Within this framework, the dynamics reflect the competition among elastic stiffness ($K$), commensurate locking ($V_c$), quenched disorder ($h_p$), external drive ($E(t)$), and an effective thermal fluctuation scale ($T_{eff}$). In the commensurate state, domain walls are pinned both by spatial disorder $h_p(r)$ and by the intrinsic lock-in potential $-V_c \cos(N\phi)$, where $N = 13$ is the commensurability index of the C-CDW phase in 1T-TaS$_2$. As a consequence, the free-energy landscape contains multiple metastable minima separated by barriers associated with local phase slips and domain-wall rearrangements.

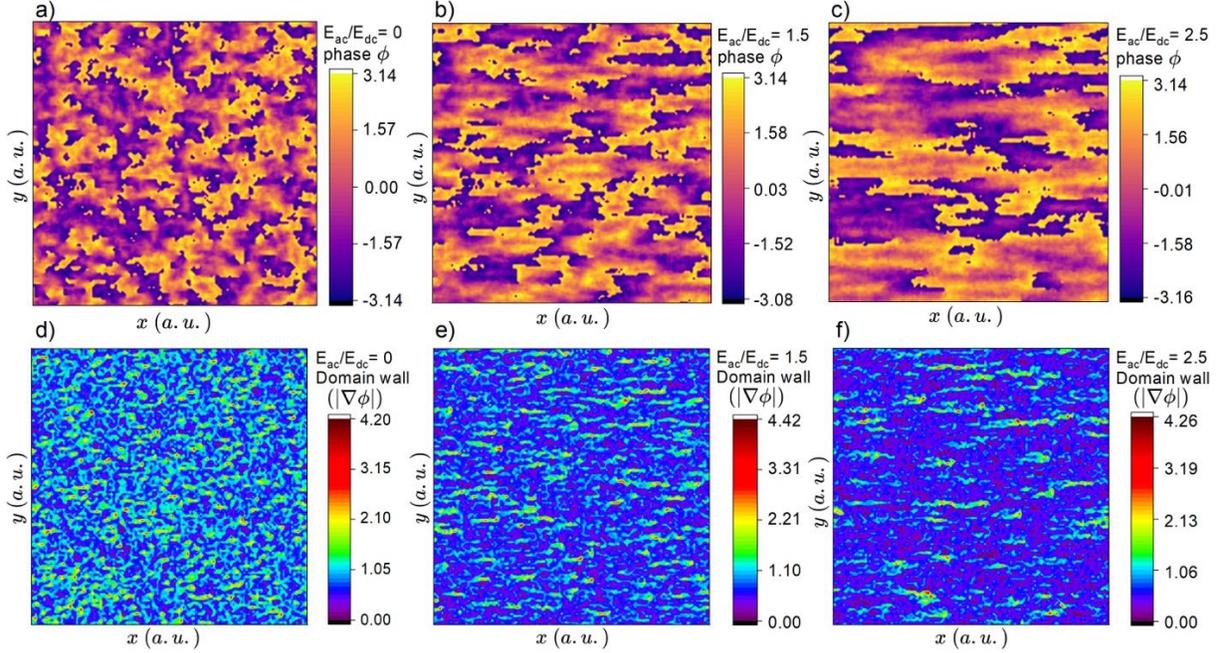

**Figure 5: The TDGL simulation of commensurate CDW domains.** Two-dimensional phase maps obtained from TDGL simulations are shown for different driving condition: (a) cooled state with $E_{ac}/E_{dc} = 0$, (b) $E_{ac}/E_{dc} = 1.5$ and (c) and $E_{ac}/E_{dc} = 2.5$. Panel (d) to (f) show the corresponding domain wall intensity under the same driving conditions. As the AC amplitude increases, the domain wall intensity decreases, and the phase maps exhibit enhanced coherence.

In Figure 5, the phase $\phi(r)$ and the magnitude of the phase gradient $|\nabla\phi|$ are plotted in (a–c) and (d-f), respectively. From left to right, the plots show snapshots after zero-field cooling (a, d), with $E_{ac}/E_{dc} = 1.5$ (b, e), and with $E_{ac}/E_{dc} = 2.5$ (c, f). The phase $\phi(r)$ provides a domain map, and the magnitude of the gradient $|\nabla\phi|$ provides a domain wall intensity map. As the $E_{ac}$ is increased, more coherent regions are formed and the domain wall intensity is reduced. Under AC–DC driving, the system periodically approaches or exceeds the local depinning threshold, allowing transient unpinning of domain segments without inducing sustained heating. In the adiabatic regime, when the drive frequency satisfies $f \ll \tau_{DW}^{-1}$, where the $\tau_{DW}$ is the intrinsic relaxation time of the domain walls, the system tracks the instantaneous force quasi-statically, enhancing barrier crossing in a manner analogous to "shake annealing." Repeated small perturbations enable escape from shallow metastable states without globally melting the ordered phase, resulting in improved phase coherence (Figure 5(c)) and a reduction in domain-wall intensity (Figure 5(f)).

To quantify AC-assisted domain-wall rearrangements, we define the escape rate from a metastable minimum, separated by energy barrier $\Delta U_0$, in the absence of drive as,

$$\Gamma_0 \sim \Omega_0 \exp\left(-\frac{\Delta U_0}{T_0}\right), \tag{1}$$

where $\Omega_0$ is the attempt frequency determined by the curvature of the free-energy landscape and $T_0$ is the base temperature. In the adiabatic limit under AC–DC driving, the period-averaged escape rate, $\bar{\Gamma}$, scales approximately as,

$$\bar{\Gamma} \propto \exp\left[-\frac{\Delta U_0}{T_0}\left(1 - \frac{E_{ac}}{E_c}\right)\right], \tag{2}$$

Where, $E_c$ is the characteristic depinning field set by pinning strength and elastic stiffness. Thus, even a modest oscillatory amplitude $E_{ac}$ can exponentially enhance barrier crossing without requiring sustained heating, enabling reorganization toward lower-energy domain configurations. This framework allows us to simulate a regime in which elasticity, commensurate locking, disorder, thermal fluctuations, and external drive all compete on comparable energy scales. Such conditions are physically relevant for describing metastable domain dynamics and AC-assisted annealing in commensurate CDW systems.

To interpret the AC–DC transport behavior of 1T-TaS$_2$ devices, we employ regime-dependent phenomenological models. In the low-frequency limit (Figure 6(a)–(d)), the response is described using a bistable switching circuit model, while in the higher-frequency regime (Figure 6(e)–(h)) we map TDGL-derived domain-wall configurations into a percolative resistor–capacitor network to capture the transport dynamics. In the low-frequency regime, the response is captured using a switching circuit model, in which the 1T-TaS$_2$ channel is represented as a bistable, state-dependent resistance that switches between a low-conductivity NC-CDW state ($g_l$) and a high-conductivity IC-CDW state ($g_h$). The values of $g_l$ = 1.28 mS and $g_h$ = 8.62 mS are selected for the representative device at $T$ = 240 K (Figure 2 (b)). In addition, a lumped external series resistance $R_s \approx 10\ \Omega$ and a parasitic capacitance $C_p \approx 2$ pF are included. Figure 6 (a) to (d) shows the evolution of the hysteresis under AC–DC sweep with $f$ = 1 MHz. Upon application of the AC drive, $V_{ac}$ = 0 to 50 mV, the hysteresis collapses, and splits into two distinct branches. In this model, the transition between the two resistive states is assumed to occur instantaneously once the internal device voltage crosses the corresponding thresholds; intrinsic switching times or order-parameter dynamics are not explicitly included. We note, however, that avalanche-like switching processes reported in quasi-1D CDW systems are associated with intrinsic finite timescales on the order of microseconds and if such dynamics are operative in 1T-TaS$_2$, they would naturally introduce an additional frequency dependence beyond the circuit-limited description [40].

To describe the high-frequency AC–DC response of the nearly commensurate CDW state in 1T-TaS$_2$, we adopt a multiscale framework that combines a TDGL description of the commensurate order parameter with a morphology-informed percolation network [54, 55]. The complex CDW field is first evolved through a cooling protocol in the presence of commensurate lock-in and quenched disorder, producing a spatially heterogeneous domain mosaic characterized by a structured phase texture. From this TDGL configuration, we extract the local magnitude of the phase gradient $|\nabla\phi|$, which serves as a quantitative measure of domain-wall density. After coarse-graining to the scale of the electrical lattice, this "domain-wall density field" defines a spatially structured resistive–capacitive (RC) landscape. Each bond of the network is modeled as a parallel RC element whose baseline resistance and capacitance interpolate between domain-like and wall-like values according to the local TDGL-derived 'wallness', i.e. regions of small phase gradient correspond to insulating commensurate domains with high resistance and moderate capacitance, whereas regions of large phase gradient—associated with domain walls and frustrated phase configurations—are assigned enhanced capacitance and reduced resistance, reflecting their increased electronic softness and polarizability.

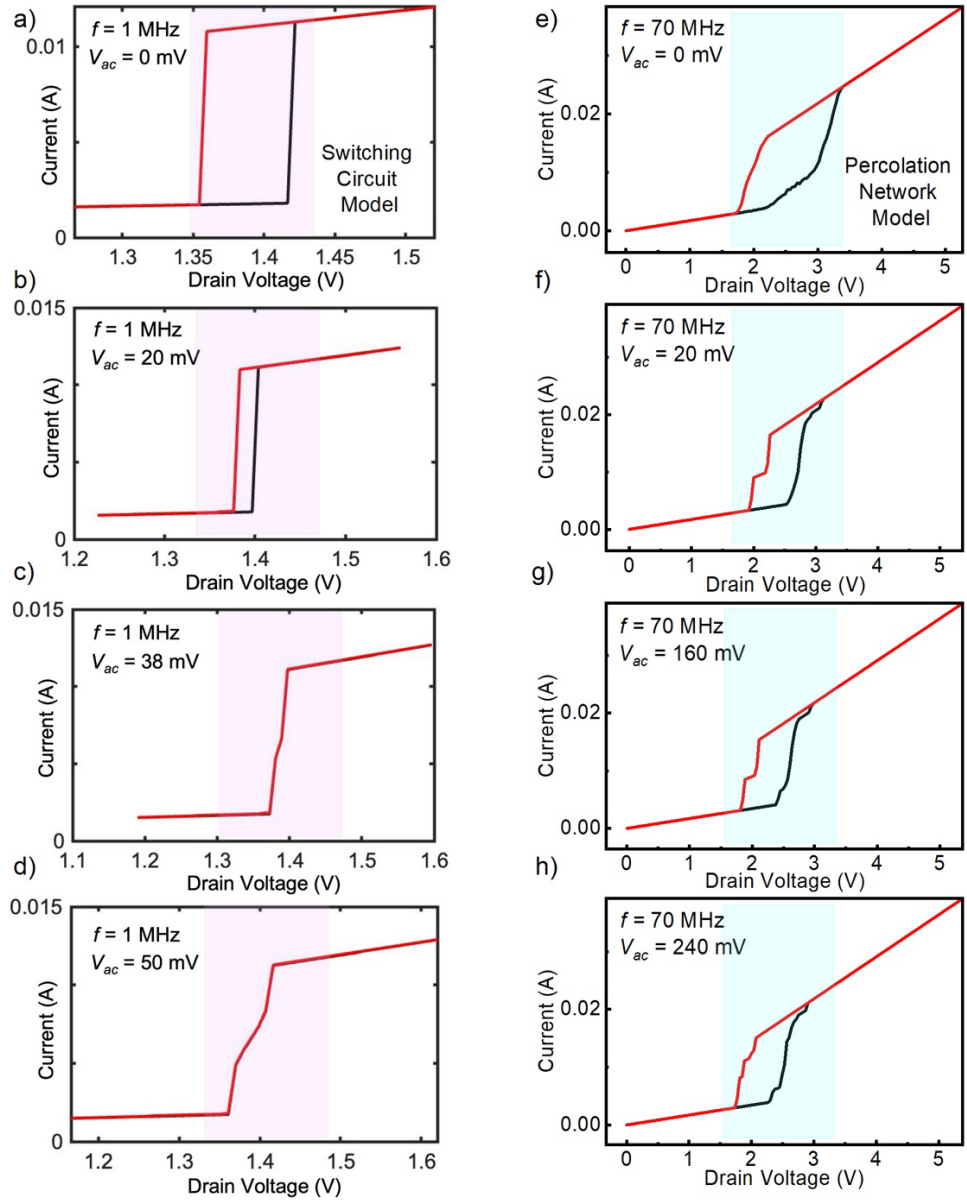

**Figure 6: AC-DC response in 1T-TaS$_2$ switching circuit and percolation-like network model.** Panels (a)–(d) capture the low-frequency response, which is described using a switching circuit model. At $f$ =1 MHz, increasing the AC amplitude from $V_{ac}$ = 0 to 50 mV leads to a collapse and splitting of the hysteresis. Panels (e)–(h) show the higher-frequency response, modeled using a TDGL–informed percolation-like resistor–capacitor network representing commensurate domains and their boundary walls. As the AC amplitude is increased, step-like features emerge in both forward and reverse sweeps.

The resulting network evolves dynamically under combined DC bias and high-frequency AC excitation. Kirchhoff's equations are solved self-consistently to determine local voltage drops, which are compared to disorder-broadened melting and nucleation thresholds governing bond switching between insulating CDW and metallic states. When the local electric field exceeds the melting threshold, bonds transition to a low-resistance metallic state; re-trapping occurs at a lower nucleation threshold, producing intrinsic hysteresis. Because the RC parameters inherit spatial correlations from the underlying CDW phase texture, switching events preferentially nucleate along pre-existing domain-wall networks, leading to morphology-guided avalanche pathways and structured percolation channels. Figure 6 (e) to (h) shows the emergence of steps in the vicinity of hysteresis, upon increasing the AC amplitude, for $f$ = 70 MHz drive frequency. This TDGL-informed percolation model thus establishes a direct link between the microscopic phase structure of the commensurate CDW and the macroscopic nonlinear AC–DC transport response, providing a physically grounded description of hysteresis and step-like features observed under high-frequency excitation.

3. **Conclusion**

In summary, RF excitation reorganizes the CDW landscape in 1T-TaS$_2$ and strongly modifies transport across the NC–IC transition regime. RF driving restructures the hysteretic transport window and generates multiple step-like features, with qualitatively distinct behavior in the low-frequency and high-frequency limits. Concurrent in-situ Raman measurements in the commensurate phase show enhanced low-frequency CDW-mode intensities together with linewidth narrowing under combined DC and RF bias, consistent with reduced dephasing, diminished inhomogeneous broadening, and increased coherence of the periodic lattice distortion. This interpretation is supported by the overdamped TDGL modeling, which captures the competition among pinning, elastic stiffness, disorder, and time-dependent driving, and shows that AC excitation can reduce domain-wall density and reorganize the discommensuration network. When this TDGL-derived mesoscale morphology is mapped onto a percolative RC transport framework with spatially varying switching thresholds and enhanced domain-wall capacitance, the model reproduces morphology-guided avalanche pathways, hysteresis restructuring, and RF-induced conductance steps. Taken together, these results support a unified physical picture in which RF fields couple directly to the mesoscale domain-wall network, reshape the underlying free-energy landscape, and thereby tune both nonlinear transport and lattice-order coherence while enabling access to metastable conduction states in quasi-two-dimensional CDW systems. More broadly, the results establish RF driving as a practical route for dynamic control of collective electron-phonon order in 1T-$TaS_2$, with implications for reconfigurable RF electronics, detectors, and mixers, as well as memory and unconventional computing architectures, including neuromorphic and nonvolatile platforms.

## 4. Experimental Section

**Device Fabrication and Electrical Testing:** Single-crystal 1T-$TaS_2$ was grown using the chemical vapor transport method. Thin flakes of 1T-$TaS_2$ with (thicknesses < 40 nm) were mechanically exfoliated onto 300 nm $SiO_2$/Si substrate. To protect the crystals from environmental degradation, thin flakes (thickness <20 nm) were encapsulated with hexagonal boron nitride (*h*-BN) using a dry-transfer technique. Electrical contacts were defined by electron-beam lithography (JEOL JSM-6610). For *h*-BN–capped devices, a dry etching step (Oxford PlasmaPro 80+) was performed to selectively remove the *h*-BN layer in the contact regions prior to metallization. Metal contacts consisting of Ti/Au (15 nm/80 nm) were deposited by electron-beam evaporation (CHA Mark-40). Temperature-dependent resistance measurements were carried out using a Quantum Design Dynacool system over a temperature range of 1.8 K– 400 K. Cryogenic AC–DC transport measurements were performed using a Lakeshore TTPX probe station in conjunction with a semiconductor device analyzer (Agilent B1500A) and a Lakeshore M81 source-measure unit. RF excitation was applied using a signal generator (Rigol DSG815) and combined with the DC bias through a bias tee (Sigatek SB12D2).

**Raman Spectroscopy:** Raman measurements were performed using a Renishaw inVia micro-Raman system with a 488 nm (blue) excitation laser. The measurements were carried out in a conventional backscattering configuration using a 50x objective. To minimize local heating, the excitation power was kept below 1 mW for all measurements. The samples were mounted in a Linkam cryostat equipped with electrical probes for bias-dependent, in-situ measurements at low temperatures.

**Time-Dependent Ginzburg–Landau Description of a Driven Commensurate CDW:** To generate a physically grounded description of the CDW domain morphology under AC–DC excitation, we simulate the two-dimensional surface of 1T-TaS$_2$ within a time-dependent Ginzburg–Landau (TDGL) framework. This model captures the competition between elastic stiffness, commensurate phase locking, quenched disorder, and time-dependent electric drive, producing spatially resolved domain-wall textures used in subsequent transport modeling. We consider a coarse-grained two-dimensional description of a commensurate charge-density wave in terms of a complex scalar order parameter $\psi(r,t) = A(r,t)e^{i\varphi(r,t)}$, where $A(r,t)$ is the CDW amplitude and $\varphi(r,t)$ is the phase. In a commensurate state, the phase is locked to discrete values $\varphi = 2\pi m/N$, where $N$ is the commensurability index. The system is described by a Ginzburg–Landau free-energy functional

$$F[\psi] = \int d^2r \left[\alpha(T)|\psi|^2 + \frac{\beta}{2}|\psi|^4 + \frac{K}{2}|\nabla\psi|^2 - V_c \cos(N\phi) - Re\{h_p h^*(r)\psi\}\right]. \quad (3)$$

The individual contributions include the Landau Term, $\alpha(T) = a(T - T_c)$, Which controls the CDW transition. For $T < T_c$, $\alpha(T) < 0$ and a finite amplitude $|\psi|$ develops. The Gradient Term, $\frac{K}{2}|\nabla\psi|^2$, penalizes spatial variations and determines the domain-wall width $\xi \sim \sqrt{K/|\alpha|}$. The term $-V_c \cos(N\phi)$ is the commensurate lock-in term that enforces discrete phase minima corresponding to the commensurate CDW. The quenched pinning is defined with $-Re\{h_p h^*(r)\psi\}$, where $h(r)$ is a complex random field representing random disorder.

The nonequilibrium evolution of the CDW is modeled by the overdamped time-dependent Ginzburg–Landau dynamics,

$$\frac{\partial \psi}{\partial t} = -\Gamma \frac{\delta F}{\delta \psi^*} - v(E,t)\partial_x \psi + \eta(r,t), \quad (4)$$

where, $\Gamma$ is the kinetic coefficient. The second term represents electric-field-driven advection of the CDW texture relative to the pinning landscape, and $\eta(r,t)$ is a complex Gaussian noise term. The functional derivative is

$$\frac{\delta F}{\delta \psi^*} = \alpha(T)\psi + \beta|\psi|^2\psi - K\nabla^2\psi - \frac{h_p}{2}h(r) + \frac{i}{2}NV_c \sin(N\phi)\frac{\psi}{|\psi|}. \qquad (5)$$

The electric field $E(t)$ enters the dynamics in two ways: (i) advection of the phase texture and (ii) a field-dependent effective temperature. The drift velocity is assumed to be proportional to the applied field, $v(E,t) = v_0 E(t)$, where $v_0$ is the advection coefficient. This term advects the phase texture, allowing domain structures to move through the quenched disorder landscape. A depinning form can also be adapted $v(E) = v_0 sgn(E) \max(|E| - E_T, 0)$, where $E_T$ is an effective threshold field. To account for field-induced fluctuations associated with Joule heating, the field-dependent effective temperature can be defined as $T_{eff}(t) = T_0 + \kappa_J E(t)^2$, where $\kappa_J$ is a phenomenological coefficient describing the strength of field-induced heating. This effective temperature controls the amplitude of the stochastic noise. The complex Gaussian noise term $\eta(r,t)$ satisfies the correlation

$$\langle \eta(r,t)\eta^*(r',t')\rangle = 2\Gamma T_{eff}(t)\delta(r-r')\delta(t-t'). \qquad (6)$$

A two-stage protocol is employed to study the evolution of domain texture under electric driving. First, the system is cooled at zero field ($E = 0$) from a high temperature $T_{High}$ to a target temperature $T_{Target}$, producing a frozen mosaic of CDW domains. After cooling, the system is allowed to relax at $T_{Target}$ and zero field to remove transient fluctuations and stabilize the domain configuration. In the second stage, the system is subjected to a combined AC–DC drive of the form,

$$E(t) = E_{dc} + E_{ac} \sin(2\pi f t), \qquad (7)$$

which can induce "shake annealing" by enabling barrier crossing without sustained overheating. The resulting evolution of the texture is characterized through several macroscopic observables, including the domain-wall density, $\rho_{DW} = \langle|\nabla\phi|\rangle$, and phase coherence, $C = \frac{|\langle\psi\rangle|}{\langle|\psi|\rangle}$, which quantify changes in domain structure and global phase ordering.

This simulation operates in a regime where elasticity, commensurate locking, disorder, thermal fluctuations, and external drive all compete on comparable energy scales, which is the physically relevant regime for metastable domain dynamics and AC-assisted annealing in commensurate CDW systems. For more information on the barrier height reduction, frequency scaling regime and parameter justification refer to the supplemental section.

**Transport Modeling Framework – Switching Circuit Model:** We used a single-element RC switching model to capture the low-frequency splitting of the hysteresis. The circuit consists of a series resistor $R_s$ feeding a bistable CDW element with two conductance states, $g_l$ (insulating) and $g_h$ (metallic), in parallel with a lumped capacitance $C_p$ to ground. The node voltage across the CDW element is driven by a DC bias $V_{dc}$ plus a small sinusoidal modulation $V_{ac}\sin(2\pi f t)$, and the CDW conductance is updated instantaneously according to threshold conditions: the device switches to the metallic state when $V_o$ exceeds a high threshold $V_h$ and back to the insulating state when $V_o$ falls below a lower threshold $V_l$, producing a voltage-controlled hysteresis window. The DC current at each bias point is obtained by time-averaging $i_s(t)$, and plotting this average current versus $V_{dc}$ for forward and reverse sweeps.

**Transport Modeling Framework – TDGL–Informed Percolative RC Network:** To compute nonlinear transport, the spatial morphology obtained from the TDGL simulation is mapped onto a coarse-grained resistor-capacitor percolation network. The final cooled TDGL configuration is used. From the TDGL snapshot, the phase field $\phi(r)$ is extracted and a scalar "wallness" field, $w(r)$, is constructed that quantifies local phase gradients. Using discrete-nearest neighbor differences, local wallness measure is defined as,

$$w(r) = \sqrt{(\Delta_x\phi)^2 + (\Delta_y\phi)^2} \ . \tag{8}$$

Where $\Delta_x\phi(r)$, and $\Delta_y\phi(r)$ is the local phase gradient along $x$ and $y$, respectively. Large values of $w(r)$ identify domain walls, where the phase changes rapidly, while small values correspond to domain interiors. The wallness map is block-averaged to a coarse $S \times S$ grid (with $S = 12$), which defines the nodes of the electrical network. Bond-resolved wallness values $w_k \in [0, 1]$ are then obtained by averaging neighboring coarse pixels and normalizing to the unit interval, where $w_k \approx 0$ corresponds to domain-interior bonds and $w_k \approx 1$ correlates with bonds located on or near phase walls. These bond wallness values constitute the only input from the TDGL morphology into the transport model.

The electrical response of this TDGL-informed morphology is modeled as a two-dimensional resistor–capacitor percolation network defined on the same $S \times S$ coarse lattice of nodes. Each bond $k$ (horizontal or vertical) is represented as a parallel combination of a resistance $R_k$ and capacitance $C_k$, whose "base" values depend on the local wallness. We interpolate linearly between domain and wall parameters,

$$R_{0,k} = R_{dom} + (R_{wall} - R_{dom})w_k, \qquad C_{0,k} = C_{dom} + (C_{wall} - C_{dom})w_k, \qquad (9)$$

such that bonds in domain interiors, $w_k \approx 0$, have $(R_{0,k}, C_{0,k}) \approx (R_{dom}, C_{dom})$ and bonds along walls, $w_k \approx 1$, have $(R_{0,k}, C_{0,k}) \approx (R_{wall}, C_{wall})$. Here, the base values before applying the local wallness and disorder, are assumed to be: $R_{dom} \approx 1000$, $R_{wall} \approx 350$, $C_{dom} \approx 10^{-14}$, $C_{wall} \approx 10^{-12}$. To mimic microscopic geometric disorder, we multiply the capacitances by a lognormal random factor $\xi_{C,k}$, drawn from $\ln(\xi_{C,k}) \sim \mathcal{N}(0, \sigma_C^2)$, so that $C_k = C_{0,k}\xi_{C,k}$. Moreover, each bond is assigned a local activation threshold, $V_{th}$, and a nucleation threshold, $V_{nuc}$, and multiplied by Gaussian disorder factors, $\xi_{\Theta,k} \sim \mathcal{N}(1, \sigma_{th}^2)$. Each bond also carries a binary state variable $s_k \in \{0,1\}$ describing whether it is in the insulating/domain-like state ($s_k = 0$) or a locally "melted" metallic state ($s_k = 1$). In the metallic state, we set the resistance to a low value $R_k = R_{metal} \ll R_{dom}$ and suppress the capacitance to a small background value, representing a highly conducting link with negligible capacitive charging (here, the metallic state value is set to be, $R_{metal} \approx 150$). Thus, the TDGL morphology affects the network through the base $(R_0, C_0)$ of each bond, while the switching threshold and metallic resistance are modeled as morphology-independent quenched disorder.

For a given DC bias $V_{DC}$ and optional RF excitation of amplitude $V_{AC}$ at a specified frequency, we solve the Kirchhoff's equations on the full lattice using a sparse complex admittance matrix. We obtain node potentials and local voltage drops on each bond for both DC, $\Delta V_k^{DC}$ and AC amplitude, $\Delta V_k^{AC}$ and construct an effective activation voltage,

$$V_k^{eff} = |\Delta V_k^{DC}| + |\Delta V_k^{AC}|. \qquad (10)$$

In addition, we include a phenomenological AC heating correction, $H_k = \Gamma_{heat} f |\Delta V_k^{AC}|^2$, which lowers the local activation threshold under strong AC drive. Whenever the effective activation voltage exceeds the local threshold, a bond in insulating state "melts" to metallic state ($s_k: 0 \to 1$). While the re-nucleation from metallic state to insulating state ($s_k: 1 \to 0$) occurs when the effective activation voltage drops below the nucleation threshold. For every DC bias value, we iterate these internal updates, re-solving the network and re-evaluating the switching conditions, until no further bonds switch. Thus, the avalanche-like collective switching can be captured.

**Acknowledgments**

The work at UCLA was supported, in part, by the Vannevar Bush Faculty Fellowship (VBFF) to A.A.B. under the Office of Naval Research (ONR) contract N00014-21-1-2947 on One-Dimensional Quantum Materials. The work at UCR and the University of Georgia was supported, in part, via the subcontracts of the ONR project N00014-21-1-2947. The nanofabrication of the test structures was performed in the California NanoSystems Institute (CNSI).

**Author Contributions**

A.A.B. conceived the idea, coordinated the project, and contributed to data analysis and manuscript preparation. R.K.L. led the theoretical development of the non-linear circuit and TDGL models and comparison with experimental data. M.T. fabricated devices, conducted transport measurements, led the experimental data analysis, assisted with the modeling, and wrote the first draft of the manuscript. Z.E.N. conducted in-situ Raman measurements. N.S. synthesized and characterized bulk crystals of 1T-TaS$_2$. T.T.S. supervised materials synthesis and characterization. T.D. contributed to circuit modeling. All authors contributed to the final draft of the manuscript.

**Supplemental Information**

The supplemental information is available on the journal website free of charge.

**The Data Availability Statement**

The data that support the findings of this study are available from the corresponding author upon reasonable request.